\documentclass[runningheads]{llncs}
\usepackage{graphicx}
\usepackage{amsmath}
\usepackage{amssymb}
\usepackage{array}
\usepackage{subcaption}
\usepackage{hyperref}
\usepackage{xcolor}
\usepackage{ulem}
\usepackage{multirow}
\usepackage{booktabs}
\usepackage{bm}
\usepackage{placeins}
\usepackage{float}
\usepackage{lineno}
\newcolumntype{C}[1]{>{\centering\arraybackslash}p{#1}}

\begin{document}
%
\title{Real World Federated Learning with a Knowledge Distilled Transformer for Cardiac CT Imaging}
\titlerunning{Real World Federated Learning for Cardiac CT Imaging}
%
\author{
    Malte T\"olle\inst{*,1,2,3} \and
    Philipp Garthe\inst{4} \and
    Clemens Scherer\inst{1,5} \and 
    Jan Moritz Seliger\inst{1,6} \and
    Andreas Leha\inst{1,7} \and
    Nina Kr\"uger\inst{1,8,9,10} \and 
    Stefan Simm\inst{1,11} \and 
    Simon Martin\inst{1,12} \and 
    Sebastian Eble\inst{2} \and
    Halvar Kelm\inst{2} \and
    Moritz Bednorz\inst{2} \and
    Florian André\inst{1,2,3} \and
    Peter Bannas\inst{1,6} \and
    Gerhard Diller\inst{4} \and
    Norbert Frey\inst{1,2,3} \and
    Stefan Groß\inst{1,11} \and
    Anja Hennemuth\inst{1,6,8,9,10} \and 
    Lars Kaderali\inst{1,11} \and
    Alexander Meyer\inst{1,8} \and
    Eike Nagel\inst{1,12} \and
    Stefan Orwat\inst{4} \and
    Moritz Seiffert\inst{13} \and
    Tim Friede\inst{1,7} \and
    Tim Seidler\inst{1,14,15} \and 
    Sandy Engelhardt\inst{1,2,3}
}
\authorrunning{M. T\"olle et al.}
%
\institute{
    DZHK (German Centre for Cardiovascular Research), all partner sites \and
    Department of Cardiology, Angiology and Pneumology, Heidelberg University Hospital, Heidelberg, Germany \and
    Informatics for Life Institute, Heidelberg, Germany \and
    Clinic for Cardiology III, University Hospital M\"unster, M\"unster, Germany \and
    Department of Medicine I, LMU University Hospital, LMU Munich, Munich, Germany \and
    Department of Diagnostic and Interventional Radiology and Nuclear Medicine, University Medical Center Hamburg-Eppendorf, Hamburg, Germany \and
    Department of Medical Statistics, University Medical Center G\"ottingen, G\"ottingen, Germany \and
    Deutsches Herzzentrum der Charité (DHZC), Institute of Computer-assisted Cardiovascular Medicine, Berlin, Germany \and
    Charité – Universit\"atsmedizin Berlin, corporate member of Freie Universit\"at Berlin and Humboldt-Universit\"at zu Berlin, Berlin, Germany \and
    Fraunhofer Institute for Digital Medicine MEVIS, Bremen, Germany \and
    Institute of Bioinformatics, University Medicine Greifswald, Greifswald, Germany \and
    Institute for Experimental and Translational Cardiovascular Imaging, Goethe University, Frankfurt am Main, Germany \and
    Department of Cardiology, University Heart and Vascular Center Hamburg, University Medical Center Hamburg-Eppendorf, Hamburg, Germany \and
    Department of Cardiology, University Medicine G\"ottingen, G\"ottingen, Germany \and
    Department of Cardiology, Campus Kerckhoff of the Justus-Liebig-University at Gießen, Kerckhoff-Clinic, Gießen, Germany\\
    \email{malte.toelle@med.uni-heidelberg.de}
}
\maketitle              
\begin{abstract}
    Federated learning is a renowned technique for utilizing decentralized data while preserving privacy. 
    However, real-world applications often face challenges like partially labeled datasets, where only a few locations have certain expert annotations, leaving large portions of unlabeled data unused. 
    Leveraging these could enhance transformer architectures’ ability in regimes with small and diversely annotated sets.
    We conduct the largest federated cardiac CT analysis to date ($n=8,104$) in a real-world setting across eight hospitals. Our two-step semi-supervised strategy distills knowledge from task-specific CNNs into a transformer. 
    First, CNNs predict on unlabeled data per label type and then the transformer learns from these predictions with label-specific heads. 
    This improves predictive accuracy and enables simultaneous learning of all partial labels across the federation, and outperforms UNet-based models in  generalizability on downstream tasks. 
    Code and model weights are made openly available for leveraging future cardiac CT analysis.
\end{abstract}

\setcounter{footnote}{0}

\section*{Introduction}

\begin{figure}[H]
    \centering
    \includegraphics[width=0.62\linewidth]{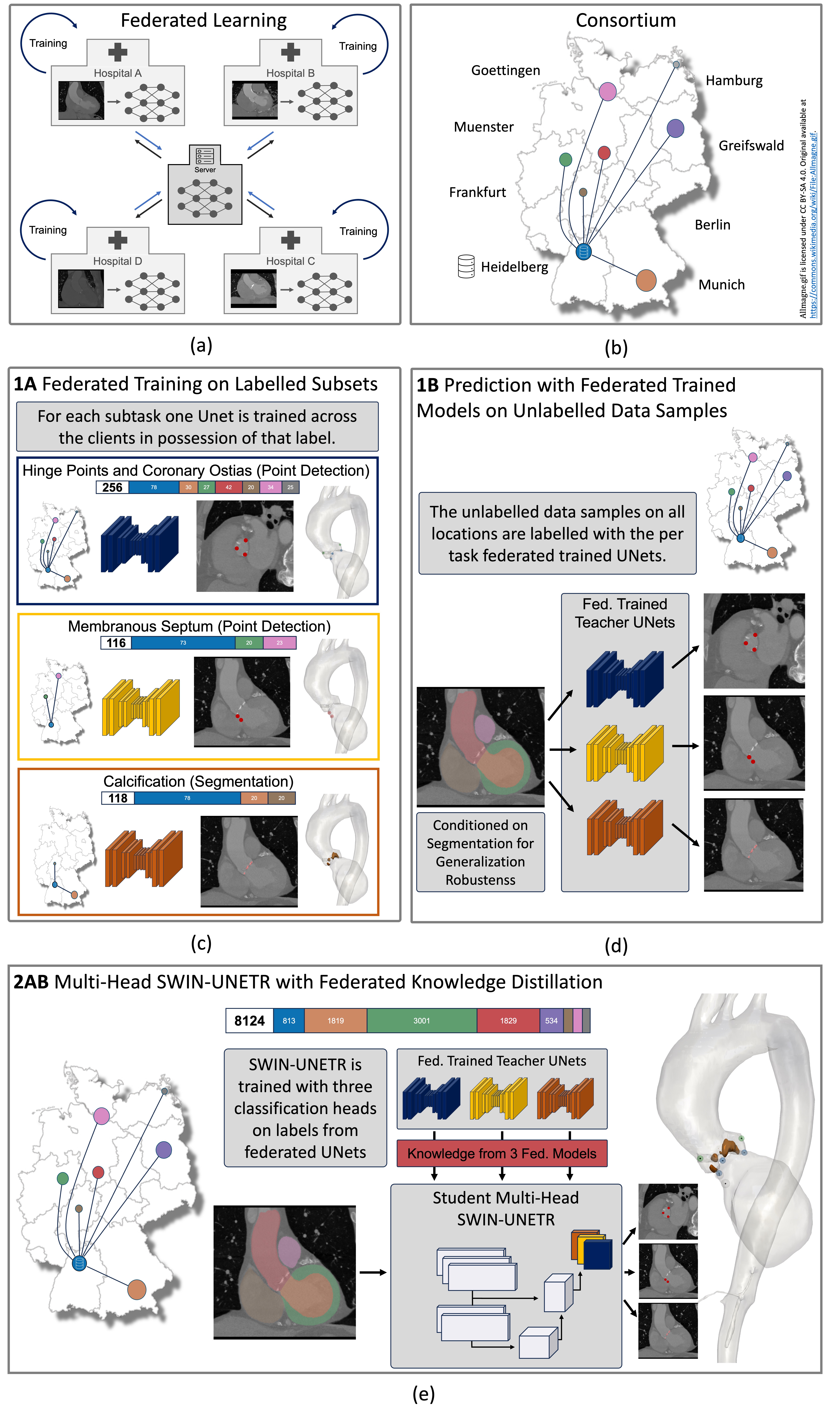}
    \caption{
        Overview of federated consortium and federated knowledge distillation (KD) training pipeline.
        a) Federated learning procedure and b) our consortium across eight university hospitals in Germany.
        c) Each label subset is not present at all locations (\textbf{Stage 1A}).
        One model (UNet) is trained for each subset in a federated manner across the locations in possession of that label.
        d) Subsequently, the federated trained models are used to make predictions on the unlabeled data samples (\textbf{Stage 1B}).
        e) The transformer based- model is trained from the predictions of the teacher network with three heads but the same backbone (\textbf{Stage 2AB}).
        Last, only the heads are fintuned on the human annotated data samples.
        Naming is consistent with Figure~\ref{fig7:method_short}.
    }
    \label{fig1:method}
\end{figure}

The manual annotation of medical images is a laborious task that requires expert knowledge~\cite{Radsch2023,Rahimi2021}.
Often, physicians can only label a limited amount of data for deep learning model training. 
They typically focus on labeling data relevant to their specific research needs, leaving a significant portion of data unlabeled and thus unused for training. 
As a result, small, highly specialized subsets of large, mostly unlabeled datasets are common in local clinics.
This presents two opportunities for improvement.
First, the training data can be enlarged by leveraging all labeled subsets across clinics, while accounting for the different structures annotated in each.
Second, by leveraging labeled and unlabeled datasets in a pooled training synergy effects can be realized, if over all participating hospitals every label of interest is present in at least one location.
Additionally, the diversity of training data from various locations can expand the overall training distribution (Figure~\ref{fig1:method}).

Privacy laws hinder the widespread collection of such heterogeneous large scale datasets stored at a single location~\cite{gdpr2016}.
Federated Learning (FL) is one paradigm that circumvents privacy concerns by reverting the paradigm of central data storage~\cite{Rieke2020,Kaissis2020,Kaissis2021,Sadilek2021}.
In FL, the model is distributed to all data holding locations, where training is performed locally before the model is sent back to a central server.
On this server the trained model weights from all participating locations are averaged before another round of training is initialized (see Figure~\ref{fig1:method}a).
Unfortunately, the quality and consistency of labels across different locations can vary, impacting the model's performance. 
Without inspection from the data scientist label quality and consistency must be ensured in FL, which often poses a big challenge that impedes the predictive performance of federated trained models on real world data~\cite{Kaissis2020}.

In situations where each hospital has a different subset of the total training labels, the locations are termed partially labeled. 
Training on such locations requires complex algorithms for handling the loss computation, where labels are not present.
Partially labeled data can further result in a skewed distribution of labels across locations. 
Some labels might be overrepresented in the overall dataset, while others are underrepresented. 
This can lead to biased models that perform well on some labels of data but poorly on others.
Training a single model to effectively address all tasks across these locations is challenging due to the uneven distribution of annotations.

The largest FL study on 3D medical images to date ($n=6,314$ patients) was performed by Pati et al.~\cite{Pati2022}, who trained an automatic tumor boundary detector for the rare disease of glioblastoma in a federated manner.
They reported improvement over a publicly trained model especially on rare cases that are not represented in rather small public datasets.
Other works include the prediction of future oxygen requirement of COVID 19 patients, the histological response to breast cancer, and the diagnosis of hypertrophic cardiomyopathy from ECG and Echocardiograms~\cite{Dayan2021,Terrail2023,Goto2022}.
The largest federated learning study in 3D cardiovascular imaging is conducted by Linardos et al.~\cite{Linardos2022}.
They use subsets of the publicly available magnetic resonance imaging (MRI) datasets from the Multi-Centre, Multi-Vendor, and Multi-Disease Cardiac Image Segmentation Challenge (M\&Ms) and Automated Cardiac Diagnosis Challenge (ACDC) with 180 patients in total~\cite{Campello2021,Bernard2018}.
In all the aforementioned studies, it is assumed that all locations possess all labels available in the federation. 
All approaches report an increase in generalizability for the federated trained model compared to the individual trained one.
To the best of our knowledge there exists no comparable study with federated learning on real world data on partially labeled datasets.
Additionally, unlabeled data is usually discarded and not used to further increase model performance.

In this work, we present a solution to train federated deep learning networks when imaging labels are scarce and their distributions are highly imbalaced over many locations. 
This presents a scenario where recent transformer architectures have severe limitations due to their dependence on large labeled cohorts~\cite{Oquab2023}.
The key contribution of this work is to use techniques from knowledge distillation to substantially increase the performance of these architectures for the purpose of leveraging their strengths towards solving related downstream tasks on the same type of images.
Transformers, with their inherent attention mechanism, benefit from a larger receptive field and the absence of inductive biases becomes advantageous in data-rich scenarios~\cite{Liu2023,Touvron2023}.

When solving imaging tasks a convolutional UNet is most often the method of choice~\cite{Isensee2021,Kim2024,Pati2022,Ulrich2023}. 
Due to the inductive bias of the convolutional operations they tend to generalize better with smaller amounts of data than transformer based architectures~\cite{Mauricio2023}.
To bridge the gap between architectures that excel in low-data regimes (e.g., convolutional networks) and those that require large datasets (e.g., transformers), knowledge distillation (KD) can be used.  
Three primary types of distillation can be distinguished: response-, feature-, and relation-based~\cite{Gou2021,Wu2022}. 
Response-based KD focuses on aligning the final output predictions (logits) of the teacher and student models~\cite{Hinton2014,Zhang2019}. 
Its architecture-agnostic nature makes it particularly suitable when teacher and student models differ in structure, as it trains the student to mimic the teacher’s output distributions for better generalization. 
Feature-based KD, on the other hand, aligns the intermediate feature representations of both models, requiring compatible architectures to effectively match features at specific layers~\cite{Passban2021,Touvron2021}. 
Finally, relation-based KD emphasizes teaching the student to capture relationships between data samples as modeled by the teacher~\cite{Chen2021,Passalis2020}.
Originally proposed for model compression, where a smaller student network learns to mimic the outputs of a larger teacher network~\cite{Hinton2014}, KD extends well beyond this use case.
All above KD approaches are at their core fundamentally a method for transferring knowledge from a teacher model to a student model, regardless of their sizes or architectures~\cite{Afonin2022}. 
Additionally, recent works on foundation models have effectively utilized KD to transfer knowledge from large pre-trained teacher models to student models, highlighting its usability in training large scale models~\cite{Oquab2023}.

Within the German Center for Cardiovascular Diseases (DZHK) we have set up a federated learning infrastructure connecting eight cardiology and radiology departments of university clinics in Germany. 
Each location provides CT scan of patients with aortic stenosis and the corresponding label types.
One caveat of dealing with real world clinical data is the heterogeneity of available labels, which is especially prominent in our use case.
While the annotated hinge points and coronary ostia are labeled across all participating locations, the membranous septum as well as the calcification are only labeled at a few not completely overlapping locations.
Furthermore, a large quantity of CT data is completely unlabeled.
Consequently, our approach includes two major factors that enlarge the data distribution used for model training: 1) the unlabeled data samples in the hospitals and 2) the federated learning approach.
More precisely, this work shows the following contributions.

\begin{itemize}
    \item Study size and label scarcity: We present the largest up to date study in cardiac computed tomography imaging from real world patient data spanning eight hospitals in Germany ($n=8,104$ images). 
    In our study, labels are scarce, meaning not all locations are in the possession of all label categories and further only a small fraction of data samples are labeled at the respective locations. 
    
    \item Federated point detection and segmentation: We train a convolutionalmodel for each custom task i.e. label type (hinge points and coronary ostia points, points of membranous septum, and segmentation of calcification) in a federated manner, which we refer to as \textbf{stage 1} of our proposed learning method~\cite{Isensee2021}. 
    Due to their inductive bias convolutional neural networks can generalize better with small amounts of training samples. 
    We show the superiority of the federated approach for each subtask.
    
    \item Semi-supervised two stage learning strategy: 
    We are the first to employ federated knowledge distillation (\textbf{stage 2}) to fuse the knowledge of the per-task models (from stage 1) into a different architecture than the teacher when small amounts of manual annotations are available.
    With the CNNs predictions are generated on the unlabeled datasets functioning as pseudo labels for training the transformer-based architecture.
    The two-stage approach increases the amount of training data mitigating the performance difference between transformer and convolutional UNet by semi-supervised learning.

    \item Downstream task: We show better generalizability of our trained transformer model compared to the convolutional based one on the downstream task of segmenting the coronary arteries by only finetuning the last layer.
    We attribute this to the learning of global context of the transformer model given sufficient data.
    
    \item Inter-observer variability: To quantify the influence of the inter-observer variability of the manual annotation on the final predictive performance every annotator in the clinics labeled samples of a public dataset~\cite{Zeng2023}.
    The inter-observer variability across locations serves as a lower-bound for the performance of the model.
    The labels of this cohort will be made publicly available.

    \item Privacy-Preserving Label Quality Visualization: Due to its privacy by design structure FL does not enable the inspection of label quality at the participating sites. To verify consistency we compare the relative location of landmarks across locations, which does not disclose patient information but allows for qualitative privacy-preserving outlier detection.
    
    \item Open source code and model weights: The code will be made publicly available. 
    Further, we release the model weights of the final transformer model, which can be used as a base model in cardiac CT imaging for future studies.
    
\end{itemize}

Our use case for the developed method focuses on improving the analysis of cardiac CT imaging for Transcatheter Aortic Valve Implantation (TAVI) patients.
Diseases of the cardiovascular system amount for up to a third of deaths in developed countries~\cite{WHO2021}. 
A common valve pathology is described by aortic valve stenosis, which is a condition where the aortic valve becomes narrowed, leading to reduced blood flow from the heart to the rest of the body. 
TAVI is a catheter-based procedure to replace the narrowed valve with an artificial one, necessitating precise imaging and analysis for optimal outcomes.
Due to its less invasive nature it has become the gold standard for treating severe aortic stenosis in patients who are considered high risk or inoperable for surgical aortic valve replacement~\cite{Genereux2012,Leon2010}. 
However, patients receiving TAVI are more prone to be dependent on a pacemaker post implantation due to the prosthesis applying pressure to the stimulation conduction system of the heart~\cite{Sammour2021}. 
Known influencing parameters are the aortic valve geometry, the per-cusp calcification, and the distance of the annulus plane to the membranous septum~\cite{Mauri2020,Musallam2022}. 
The three hinge points define the location of the aortic annulus plane, which is the location of the smallest diameter of the aortic root and, thus, determines the size of the prosthesis, while the coronary ostia determine the possible length. 
A measurement not yet taken in clinical practice in an automatic way is the location of the smallest part of the membranous septum and its distance from the annulus plane~\cite{Jorgensen2022}. 
Multiple works exist that perform localization of aortic root and hinge points as well as coronary ostia~\cite{Aoyama2022,Astudillo2019,Kruger2022}. 
All methods were trained on single-site data, lacking the ability to quantify all CT aspects due to missing labels for certain subtasks. Consequently, these approaches may not generalize well beyond their training datasets. No existing method combines aortic landmark detection with membranous septum detection and aortic root calcification quantification, which are key predictors for prosthesis selection and TAVI outcomes.

\section*{Results}

The results are presented as follows. First, we describe how federated training enhances performance. 
Second, we present the effects of our proposed two-stage learning procedure. 
Next, we assess the consistency and reliability among labelers in a privacy-preserving manner by evaluating label quality using known anatomical relationships.
This leads us to examine the impact of inter-observer variability on model predictions, a critical issue in federated learning. 
Finally, we evaluate our model’s generalization performance on a public dataset for a different task.

\subsection*{(Semi-supervised) Federated Knowledge Distillation from Partially Labeled Datasets}

For evaluation purposes we perform a large series of experiments comparing different architectures and local vs. federated training. 
For each task (point detection of hinge points, coronary ostia, membranous septum, and segmentation of calcification) three different methods are compared with different models.
We train a convolutional UNet as well as two transformer architecture (ViT for segmentation and SWIN-UNETR)~\cite{Hatamizadeh2022a,Hatamizadeh2022b}.
While ViT is based on conventional self-attention, SWIN-UNETR employs a shifted window attention approach that trades of global and local context.
First, we train a UNet and the two transformer-based models on each local dataset.
Second, all models are trained in a federated fashion across the locations having these labels.
Third, we perform semi-supervised federated knowledge distillation on the unlabeled data of each hospital with our federated trained UNet as the teacher and a ViT as well as a SWIN-UNETR as student, before finetuning on the labeled subsets.
For federated training we always leave at least one location out for training for having an independent testset in form of a completely separated dataset.

\begin{figure}[]
    \centering
    \includegraphics[width=0.8\linewidth]{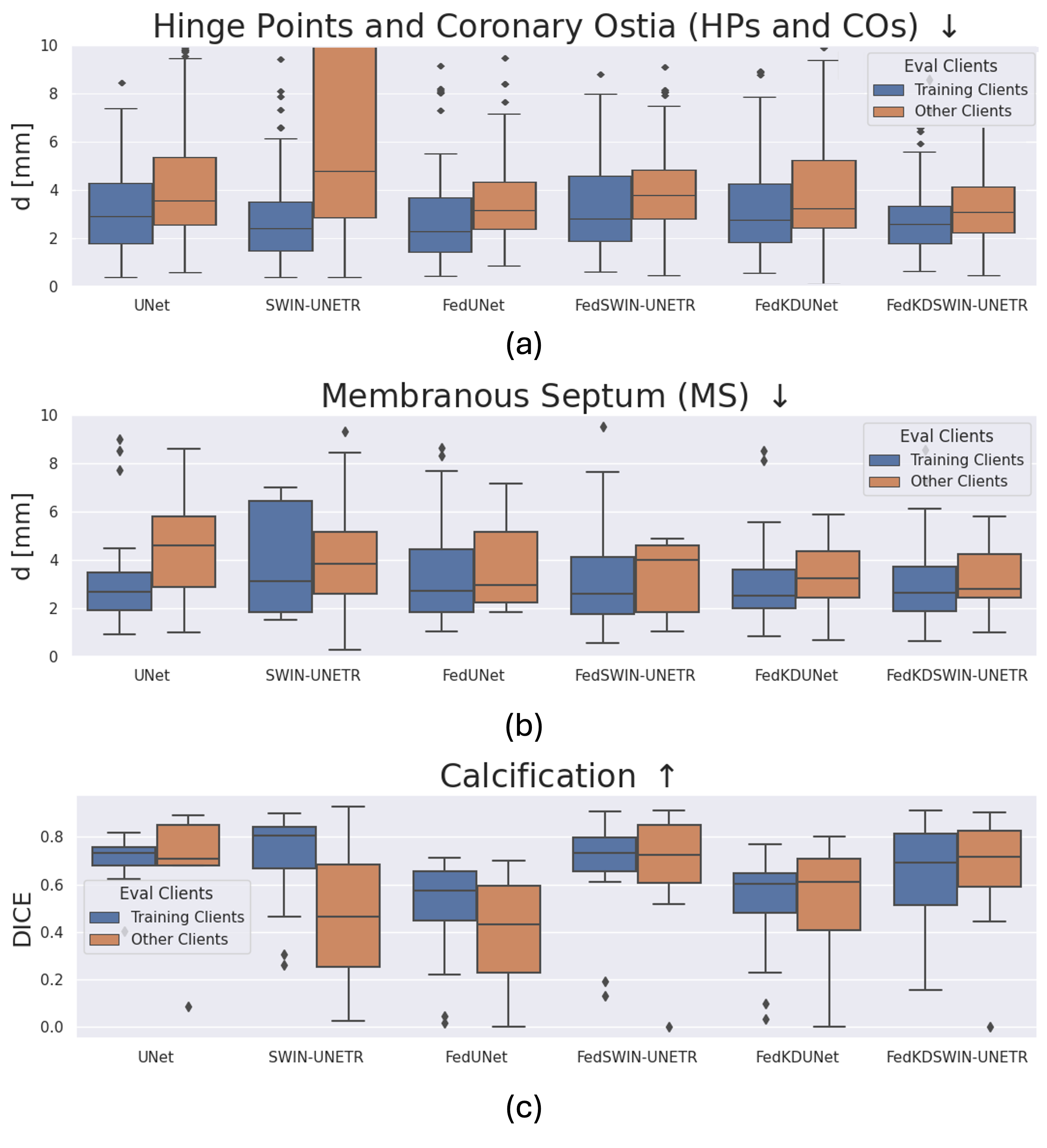}
    \caption{
        Comparison of UNets and transformer-based model (SWIN-UNETR) in boxplots for local, federated, and federated KD training for a) Hinge Points \& Coronary Arteries (HPS \& CAs), b) Memebranous Septum (MS), and c) Calcification.
        Test results on training clients are shown in blue, the results on independent test clients is shown in orange.
        In the boxplots median, 25th and 75th quartile, as well as outliers are shown.
        The locally trained models perform well on their locations's respective data, but do not generalize to the data from other locations.
        The transformer-based architecture performs worse than the Unet.
        The generalization performance can be enhanced with federated training, but the UNet still performs and generalizes better.
        After performing federated KD and subsequent finetuning the performance of the transformer-based model is on par with the UNet on detecting the hinge points, coronary ostia, and membranous septum, while outperforming it on segmenting the calcification.
        While the predictive performance of the SWIN-UNETR can be enhanced with more training samples due to KD to be better or on par with the UNet architecture, KD does not enhance the performance of the UNet to a similar degree.
    }
    \label{fig2:local-fed-kd}
\end{figure}

\begin{figure}[t]
    \centering
    \includegraphics[width=\textwidth]{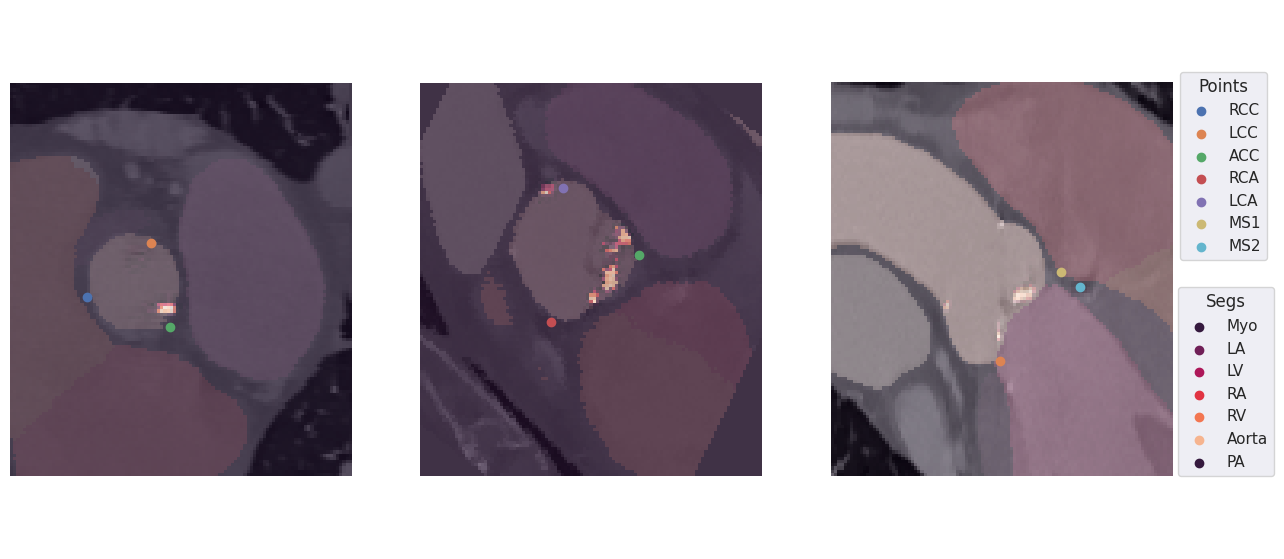}
    \caption{
        Qualitative results of the predicted labels of FedKD SWIN-UNETR.
        The predictions of our final distilled model were inspected by two experienced cardiologists verifying that the points are placed within the anatomical variance present. 
        RCC: right coronary cusp, LCC: left coronary cusp, NCC: non-coronary cusp, RCO: right coronary ostium, LCO: left coronary ostium, MS1: upper, and MS2: lower point of membranous septum, Myo: myocardium, LA: left atrium, LV: left ventricle, RA: right atrium, RV: right ventricle, PA: pulmonary artery.
    }
    \label{fig3:pred-ct-slices}
\end{figure}

The results are presented quantitatively in Table~\ref{tab:results_all} (mean and standard deviation) and their distribution in Figure~\ref{fig2:local-fed-kd} (boxplot with median and quartiles) and qualitatively in Figure~\ref{fig3:pred-ct-slices}.
Due to the obvious underperformance of ViT we do not present the results in Figure~\ref{fig2:local-fed-kd} to remain a direct comparison of the UNet and our best performing transformer-based architecture (SWIN-UNETR).
The results for all tasks and models per location can be found in the supplementary information.

As can be seen in Figure~\ref{fig2:local-fed-kd}a models trained only on the local data shards underperform on datasets from other locations.
Transformer based architectures generalize worse than convolutional UNet based ones, which we attribute to the inherent inductive bias of these architectures.
The mean distance of the predicted hinge points of the local UNet approach is at $3.09 \pm 1.71\,\mathrm{mm}$ for the same location and at $3.80 \pm 2.02\,\mathrm{mm}$ for held out test locations, while the SWIN-UNETR predicts points at a mean distance of $2.66 \pm 1.79\,\mathrm{mm}$ and $4.89 \pm 4.08\,\mathrm{mm}$ respectively.
The ViT-based model overfits the training data significantly so that it even predicts points far off for the test sets on training clients ($18.43 \pm 20.51\,\mathrm{mm}$ and $17.71 \pm 19.42\,\mathrm{mm}$).
Federated training improves generalization performance for both methods.
However, the UNet ($2.59 \pm 1.76\,\mathrm{mm}$, $3.43 \pm 1.79\,\mathrm{mm}$) performs better than the SWIN-UNETR ($3.06 \pm 1.70\,\mathrm{mm}$, $3.89 \pm 1.91\,\mathrm{mm}$).
While ViT can be improved with federated training its performance still falls short of the other two models ($5.97 \pm 7.73\,\mathrm{mm}$, $6.32 \pm 6.27\,\mathrm{mm}$).
While the performance of the SWIN-UNETR can be enhanced by performing semi-supervised federated knowledge distillation from the federated trained UNet on the previously unlabeled data samples at all locations the performance of the KD UNet is similar to the federated one.
The predicted points lie at a mean distance of $2.80 \pm 1.71\,\mathrm{mm}$ for the training locations and $3.36 \pm 1.83\,\mathrm{mm}$ for the held out test locations for the transformer and at $3.18 \pm 1.92\,\mathrm{mm}$ and $3.83 \pm 2.12\,\mathrm{mm}$ for the UNet.
The performance for ViT can also be improved with our two-stage federated learning strategy, but the results again fall short ($5.76 \pm 3.25\,\mathrm{mm}$, $5.81 \pm 4.32\,\mathrm{mm}$).

{
\renewcommand{\arraystretch}{1.3}
\begin{table}[]
    
    \centering
    \caption{
    Results of model architectures for the different learning schemes and label types.
    Three model types are investigated: convolutional UNet, a vision transformer (ViT) for segmentation, and SWIN-UNETR, which uses a siding attention approach different to a conventional ViT.
    All architectures are trained locally per location (Local), federated across labeled subsets (Fed), and with our federated knowledge distillation (FedKD) approach.
    Results are reported for locations the model was trained at (Training) and tested at the remaining (Other).
    All values are presented as $\mathrm{mean} \pm \mathrm{std}$.
    }
    \label{tab:results_all}
    \begin{tabular}{p{2.4cm}|p{1.3cm}|>{\centering\arraybackslash}p{1.8cm}|>{\centering\arraybackslash}p{1.8cm}|>{\centering\arraybackslash}p{1.8cm}|>{\centering\arraybackslash}p{1.8cm}|>{\centering\arraybackslash}p{2.1cm}|>{\centering\arraybackslash}p{2.1cm}}
        \multirow{2}{*}{Training Scheme} & \multirow{2}{*}{Model} & \multicolumn{2}{c|}{HPs \& COs $\downarrow$ [mm]} & \multicolumn{2}{c|}{MS $\downarrow$ [mm]} & \multicolumn{2}{c}{Calc $\uparrow$ [DICE]} \\
        \cline{3-8}
         & & Training & Other & Training & Other & Training & Other \\
         \hline
         \multirow{3}{*}{Local} & UNet & $3.48 \pm 2.77$ & $4.27 \pm 2.94$ & $3.01 \pm 1.84$ & $4.30 \pm 1.82$ & $0.708 \pm 0.103$ & $0.644 \pm 0.290$ \\
         & ViT & $9.45 \pm 11.87$	& $14.85 \pm 16.35$ & $6.86 \pm 11.14$ & $37.88 \pm 33.10$ & $0.644 \pm 0.184$ & $0.474 \pm 0.275$ \\
         & SWIN & $2.66 \pm 1.79$ & $4.89 \pm 4.08$ & $3.96 \pm 2.19$ & $4.06 \pm 2.16$ & $0.709 \pm 0.190$ & $0.466 \pm 0.265$ \\
         \hline
         \multirow{3}{*}{Fed} & UNet & $2.91 \pm 2.54$ & $3.75 \pm 2.38$ & $3.27 \pm 2.02$ & $3.75 \pm 1.96$ & $0.495 \pm 0.209$ & $0.391 \pm 0.212$ \\
         & ViT & $4.75 \pm 4,17$ & $3.71 \pm 1.88$ & $3.82 \pm 2.50$ & $5.32 \pm 4.98$ & $0.671 \pm 0.191$	& $0.636 \pm 0.285$ \\
         & SWIN & $3.53 \pm 2.82$ & $3.98 \pm 2.05$ & $3.03 \pm 1.85$ & $3.30 \pm 1.60$ & \textbf{\bm{$0.683 \pm 0.202$}} & \textbf{\bm{$0.692 \pm 0.232$}} \\
         \hline
         \multirow{3}{*}{FedKD} & UNet & $3.54 \pm 2.85$ & $4.25 \pm 2.94$ & $2.99 \pm 1.81$ & $3.40 \pm 1.56$ & $0.527 \pm 0.209$ & $0.526 \pm 0.228$ \\
         & ViT & $4.70 \pm 4.14$ & $3.72 \pm 1.88$ & $3.28 \pm 2.31$ & $4.35 \pm 2.34$ & $0.562 \pm 0.200$ & $0.566 \pm 0.240$ \\
         & SWIN & \textbf{\bm{$3.04 \pm 2.34$}} & \textbf{\bm{$3.54 \pm 2.12$}} & \textbf{\bm{$2.95 \pm 1.72$}} & \textbf{\bm{$3.29 \pm 1.45$}} & $0.646 \pm 0.208$ & $0.670 \pm 0.231$ \\
    \end{tabular}
    \scriptsize{HPs \& COs = Hinge Points and Coronary Ostia, MS = Membranous Septum, Calc = Calcification, KD = Knowledge Distillation}
\end{table}
}

The performance for detecting the membranous septum is similar to localizing the hinge points.
The UNet generalizes better with fewer data samples, but the SWIN-UNETR can be improved with semi-supervised federated knowledge distillation to even surpass the UNet on the unseen test locations.
The local UNet predicts a mean distance of $3.01 \pm 1.84\,\mathrm{mm}$ on the same client and $4.30 \pm 1.82\,\mathrm{mm}$ on others, the local SWIN-UNETR predicts points at $3.96 \pm 2.19\,\mathrm{mm}$ and $4.06 \pm 2.16\,\mathrm{mm}$ distance respectively.
The conventional ViT again overfits drastically to the training data distributions predicting points at $6.86 \pm 11.14\,\mathrm{mm}$ for the testsets of the training clients and $37.88 \pm 33.10\,\mathrm{mm}$
When training both models in a federated manner the SWIN-UNETR generalizes better when knowledge distillation is employed, what is seen with the lower standard deviation.
The UNet's mean distance lies at $3.40 \pm 1.56\,\mathrm{mm}$, while the SWIN-UNETR's is at $3.29 \pm 1.45\,\mathrm{mm}$.
The performance on the training locations is very similar (UNet: $2.99 \pm 1.81\,\mathrm{mm}$, KDT: $2.95 \pm 1.72\,\mathrm{mm}$).
While the ViT consistently underperforms both models we can see a strong improvement by employing our two-stage learning strategy.

Segmenting the calcification in the aortic root leads to different results than the previous tasks.
Both transformers perform better than the UNet especially when trained in a federated manner.
The SWIN-UNETR reaches a DICE score of $0.683 \pm 0.201$ on the testsets of the training locations and $0.692 \pm 0.232$ on the held out test location, the ViT $0.671 \pm 0.191$ and $0.636 \pm 0.285$ respectively.
The federated UNet only achieves a DICE score of $0.410 \pm 0.209$ and $0.391 \pm 0.212$.
The model trained with KD is almost on par with the federated trained SWIN-UNETR with DICE scores of $0.646 \pm 0.208$ and $0.670 \pm 0.231$.
We attribute the slightly worse performance to the concurrent point detection which seems to favor partly other image features than calcification segmentation.
The ViT's performance again falls short of the SWIN-UNETR, it achieves DICE scores of $0.562 \pm 0.200$ and $0.566 \pm 0.240$ respectively.

In conclusion, semi-supervised federated knowledge distillation enhances the predictive performance of a transformer based architecture (SWIN-UNETR)~\cite{Hatamizadeh2022b} to be better or to keep astride with the UNet based counter part.
Further, the tasks of locating the hinge points, coronary ostia, and membranous septum as well as segmenting the calcification of the aortic root can be solved with one model despite the distributed label classes across different classes.
While the two-stage learning strategy improves performance also for other transformer architectures such as the conventional ViT, the performance when employing a shifted window self-attention is better.
As we have shown that SWIN-UNETR outperforms the ViT consistently, we will focus on that architecture in the following.

\subsection*{Anatomical Relations for Visual Assessment Label Quality}

\begin{figure}[]
    \centering
    \includegraphics[width=\linewidth]{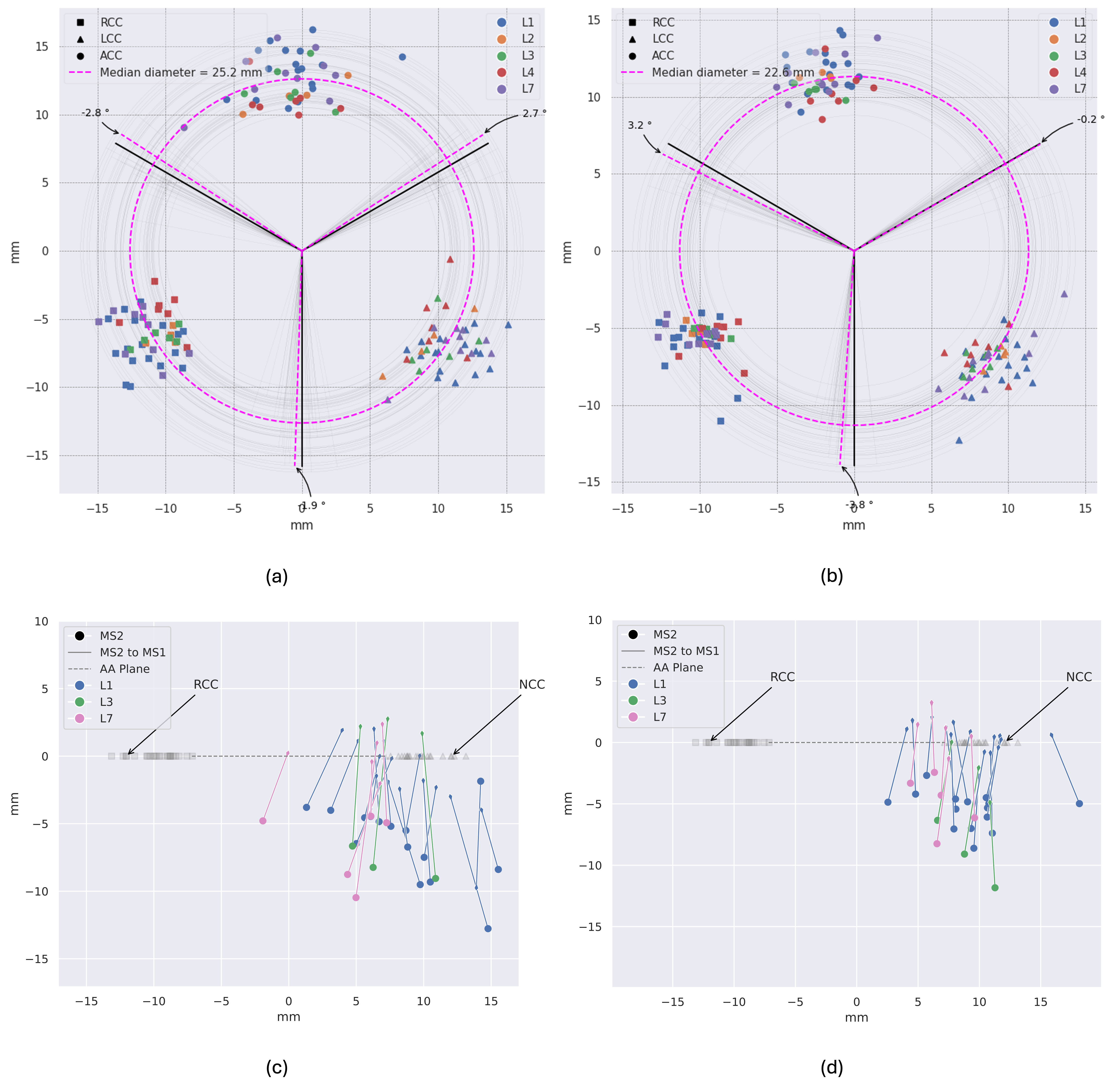}
    \caption{
        Privacy-preserving inspection of labels.
        The overall distribution of landmarks should be similar across locations, because the geometrical relations between the points is relatively homogeneous.
        a) human annotated and b) model predicted hinge points, c) human annotated and d)  model predicted membranous septum landmarks.
        In a) and b) the AA plane is defined from the three hinge points, the center point is registered, and the rotational angle is minimized to the distance from an optimal orientation of 120° between the three points.
        In c) and d) the RCC and NCC hinge points are registered and the location of the two points representing the membranous septum in relation to the two points is visualized.
        Thus, the overall quality of labels without disclosing any image information can be inspected.
        In c) MS1 and MS2 are confused (arrow points down).
        The spread is larger for the human annotated labels, which we attribute to slightly different annotation habits.
        RCC: right coronary cusp, LCC: left coronary cusp, NCC: non-coronary cusp, MS1: upper point of membranous septum, MS2: lower point of membranous septum.
    }
    \label{fig4:anatomical-constraints}
\end{figure}

One crucial aspect that hinders the widespread usage of federated learning to data is the inferior label quality sometimes present at participating locations.
In the centralized setting one can identify   false labels from training with visual inspection.
Due to its inherent privacy constraints original data such as images cannot be shared and inspected, however, their annotations can be exchanged. We therefore compared the geometric relation of labels to each other across participating locations to find outliers or a systematic bias. For example, we identified a mix up of label ids for upper and lower membranous septum point, shown in Figure~\ref{fig4:anatomical-constraints}c. 

Figure~\ref{fig4:anatomical-constraints} shows some outliers for hinge points and membranous septum.
Interestingly, the spread of labels is larger in the manual annotations, while the predictions of the network are more centered. 
This indicates a higher inter-observer variability, which we separately assessed in the following section. 
Furthermore, no confusion of point ids occurred in the predicted landmarks.

\subsection*{Evaluation of Inter-Observer Variability on Public Dataset}

\begin{figure}[t]
    \centering
    \includegraphics[width=\linewidth]{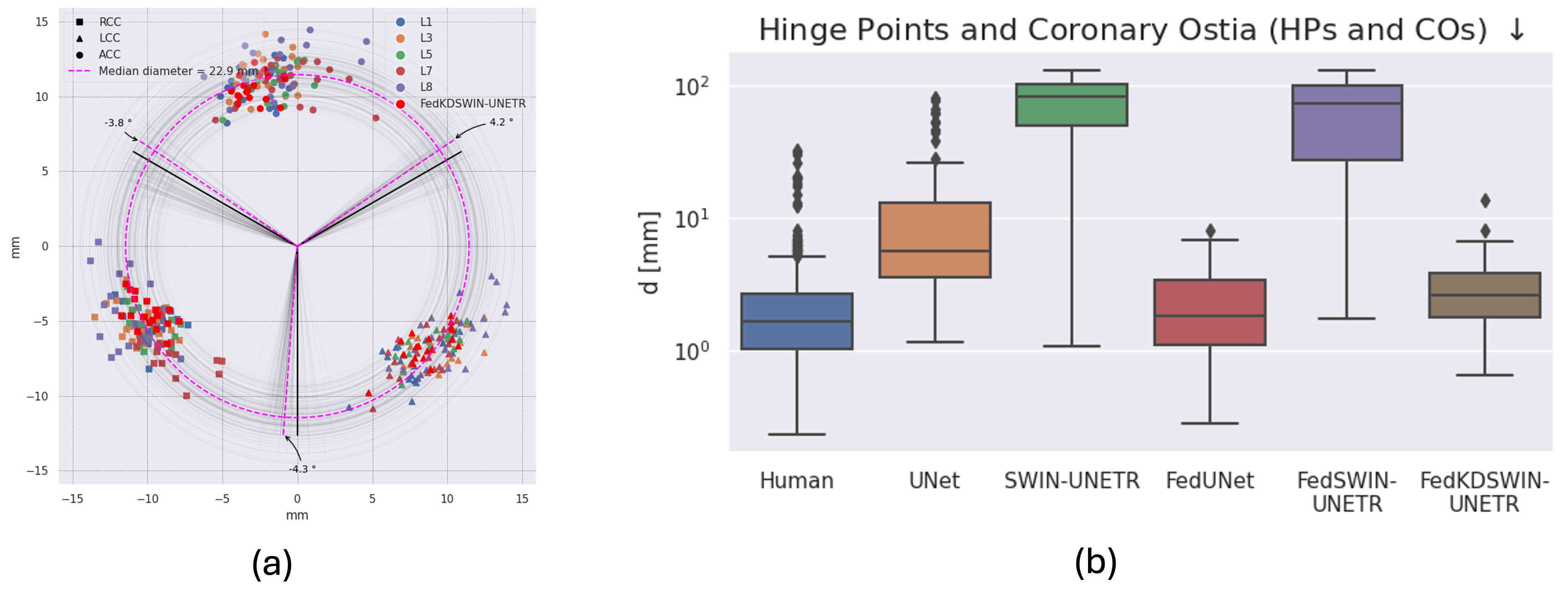}
    \caption{ 
        Distribution of hinge point labels across annotators and boxplot of distance of annotators and trained models on public dataset.
        Labels were obtained for the public dataset ImageCAS~\cite{Zeng2023}, which also serves as out-of-distribution testset.
        a) All annotators have placed the hinge points at the correct location. However, some systematic differences can be observed (e.g. between RCC of location 2 and 4). 
        b) Average distance from mean points in boxplot (median as well as 25th and 75th quartile). 
        The median distance for the human annotators is around 2\,mm. Convolutional networks generalize better for local (on one location only) and federated training. 
        By using KD on a large dataset, the performance and generalizability of transformers can be significantly enhanced. 
    }
    \label{fig5:results-public-dataset}
\end{figure}

To quantify the inter-observer variability of the manual generated ground truth to our model, we evaluated the performance of our final model on the public ImageCAS dataset~\cite{Zeng2023} against  each annotator from the participating locations, each of whom labeled 20 samples.
The mean distance from the mean over all annotations is $2.60 \pm 3.58\,\mathrm{mm}$.
Using the same method for displaying the distribution of labels as in Figure~\ref{fig4:anatomical-constraints}a and \ref{fig4:anatomical-constraints}c  the differences between human annotators from different hospitals are qualitatively explored.
Despite providing a unified annotation protocol before labelling, some systematic biases can be found, e.g., between location 2 and 4 on the hinge point of the right coronary cusp (c.f. Figure~\ref{fig5:results-public-dataset}a).
For evaluation of the trained models the 2\,mm pose a lower bound for the test error and our results show that our model is almost on-par (Figure~\ref{fig5:results-public-dataset}b).

\subsection*{Quantitative Evaluation on Public Dataset}

Since the ImageCAS dataset~\cite{Zeng2023} was not captured for TAVI patients but for inspecting the coronary arteries, a slightly different CT protocol was used.
The dataset serves as an out-of-distribution validation set to verify the generalization performance of the different methods.
The inter-observer variability has a mean of $2.60 \pm 3.58\,\mathrm{mm}$, which is the lower bound the methods can reach on average.
As was seen from the federated experiments the UNet based architectures can generalize better with less data samples (UNet: $15.54 \pm 19.02\,\mathrm{mm}$, SWIN-UNETR: $74.99 \pm 35.74\,\mathrm{mm}$).
The performance of the SWIN-UNETR degrades significantly indicating overfitting.
While the federated approach improves the performance of the UNet, the transformer is not improved in a meaningful way (FedUNet: $2.47 \pm 1.69\,\mathrm{mm}$, FedSWIN-UNETR: $74.36 \pm 33.14\,\mathrm{mm}$).
However, if semi-supervised federated KD is used to pre-train the SWIN-UNETR on the large unlabeled datasets, the performance can be increased and is in range of the federated UNet approach (FedKDSWIN-UNETR: $2.84 \pm 1.65\,\mathrm{mm}$).

\subsection*{Generalizability to Downstream Task}

Besides planning of TAVI procedures, pre-procedural exclusion of relevant coronary artery disease is recommended in these patients by current guidelines~\cite{Renker2024}. 
To investigate the generalization performance of our trained models, we opt for segmenting the coronary arteries in the public ImageCAS dataset~\cite{Zeng2023}, which already includes contours of the vessel lumen for 1000 patients (80/20 train-test-split). 

For segmenting the coronary arteries, we restrict ourselves to only finetune the last output layer of both models trained with KD, the UNet and SWIN-UNETR, to assess the extent of feature extraction already achieved by the backbone of the federated model.
In both models the last layer is a $1 \times 1$ convolution that only reweighs the feature maps from the previous layer.
While the SWIN-UNETR yields a DICE score of 0.245 the UNet is only able to achieve a DICE score of 0.045.
We attribute this to the learning of global context in the transformer encoder that enables better performances compared to convolutional based ones. 

\section*{Discussion}

We performed the largest federated learning 3D cardiac CT imaging study to date on 8104 scans across eight hospitals in Germany.
We are the first to solve the problem of federated learning on partially labeled datasets in the realm of real world medical data instead of carefully curated public challenge datasets.
In addition to training on labeled subsets of the data we also leverage the unlabeled images to increase the performance with semi-supervised federated knowledge distillation from a UNet teacher model to a transformer student model (SWIN-UNETR and ViT based)~\cite{Hatamizadeh2022a,Hatamizadeh2022b}.
The predictions of the federated trained submodels are better on the other locations compared to the single models trained on each location independently.
Surprisingly, the federated model often performs better than the own local trained one.
We attribute this to the better generalization ability of the federated model since our annotated training subsets are sometimes quite small and exhibit inter-observer variability.
The federated workflow is especially beneficial for these locations that do not possess large quantities of (labeled) data.
Our distilled SWIN-UNETR can serve as a base model for future work on cardiac CT imaging.
Moreover, we have shown its generalizability for out of distribution samples on a publicly available dataset (see Figure~\ref{fig5:results-public-dataset}).
While both transformer-based architectures achive better performance with larger dataset sizes, the performance is better for the SWIN-UNETR, which uses shifted window self-attention.
While ViT's performance is also enhanced, it's performance falls short.
We leave it for future work to examine whether this difference might potentially be mitigated with even larger dataset sizes.

The advantage of using a transformer-based model is only evident when the dataset sizes are large enough and federated training might be one ingredient to have access to many distributed data sets. However, in a setting without the presence of many human annotated samples, training transformer architectures to reach very good performance is still extremely challenging.
Our two stage approach using semi-supervised knowledge distillation with a UNet teacher model seems to be one solution to this problem.  
When training on downstream tasks the features extracted from the SWIN-UNETR seem to be more meaningful as it performs better when only finetuning the last layer, a $1 \times 1$ convolution posing a reweighting of the previous layer.

Compared to other federated learning studies our work is of higher complexity~\cite{Dayan2021,Linardos2022,Pati2021} due to different field of views and anisotropic spacing. 
Contrary to past studies where all labels for all tasks existed at all locations we deal with partially labeled ones that have a skewed distribution of present labels. 
Approaches to learning from partially labeled datasets in a federated environment include learning one encoder per participating client and label~\cite{Xu2023}.
However, this is only possible if each client is in possession of only one label.
Further, marginal loss~\cite{Shia2020} is a popular method for dealing with partially labeled datasets~\cite{Liu2023,Wang2023}.
The homogeneous distribution of anatomical structures in the human body can also be utilized in the training process to make assumptions about missing labels~\cite{Zhou2019}.
But the works are performed on large, relatively easy to segment structures (e.g. large organs such as liver).
Different classification heads for each dataset in the training distribution also represent one way of dealing with partially labeled datasets~\cite{Ulrich2023}.
However, this discards information from possibly intersecting labels across the datasets~\cite{Toelle2024b}.

Further, we are the first to employ knowledge distillation in a federated environment on real world data CT cardiac imaging data.
Our final model that is distilled from three teacher models can perform the tasks of point detection and segmentation simultaneously.
The problem solved in this work requires expert physician knowledge in contrast to solving a problem that only has a binary discriminative outcome that can be read out from a electronic health record database.
The research on federated knowledge distillation (KD) shows similarities, as these studies are conducted using publicly available datasets~\cite{Kim2024}. 
In KD, the predictive ability of a low-capacity student network is enhanced by training it to align its predictions with those of a high-capacity teacher network~\cite{Hinton2014}. 
Typically, knowledge is distilled from a group of teacher networks in FL, each trained on data from a different location~\cite{Seo2022,Mora2022}. 
Other methods include distilling knowledge by matching attention maps between client models or aligning the feature maps of both models~\cite{Wu2022,Gong2021,Tung2019}.
Wang et al. use marginal loss together with KD to learn a model across partially labeled datasets~\cite{Wang2023}.
However, marginal loss was sufficient to learn all structures present in the federation and KD was employed to further enhance results.
As stated, this is only the case when trained on large labels that are relatively easy-to-segment.

Before being able to train a model successful in a federation many tedious and practical obstacles needed to be solved.
We were only allowed to initialize communication from within the clinic networks. 
Further, we had to take additional security measures in the form of transport layer security (TLS) and username and password authentication.
We thus chose fedbiomed as library for federated learning, as they support many securtiy features out of the box~\cite{Silva2020}.
We hope that the preprocessing and training scripts for this study can be used to accelerate further studies in the future.

Once each location had successfully applied for the ethics agreement, the downloading of data from the PACS and other clinical information systems could be initiated.
Although the system is standardized even the intra-hospital variance of data was large so that site specific pre-processing was necessary.
Each hospital had different preferences regarding the recorded field of view and spacing. 
Different naming schemes made it difficult to extract the right series for each patient.
Despite all the obstacles we believe one reason why our distilled model pretrained on the unlabeled data performs better is the large data heterogeneity induced by some of above factors.

In addition to homogenizing data formats also the hardware and software used needed to be uniform.
Each location purchased the same machine to perform the learning process.
However, different requirements at each location made different installation and network specifications necessary dependent on the individual site.
As unified software solution we opted for an adapted version of Kaapana~\cite{Scherer2020}. 
It allows for flexible deployment of containerized applications.
After pseudonymization or anonymization dependent on the requirements at the individual locations the data was uploaded in the integrated PACS of our platform.
From there it could be exported, filtered and made available for federated training in a consistent manner across all locations.
Setting up the software and hardware stack required numerous conference calls~\cite{Toelle2024a}.

Federated learning has a privacy by design structure since no data leaves the individual hospitals.
However, some works have proven that in a dishonest environment clients can either corrupt the training process or reconstruct part of model's training data from the weights~\cite{Usynin2021}.
Multiple attack vectors exist that can mostly be divided in privacy- and utility-centered attacks.
Privacy-centered attacks describe the obtaining of information through unintentional leakages during training.
One such attack is Model Inversion, in which an attacker might obtain data, which was used for training, by studying how specific inputs affect the model's output~\cite{He2019,Zhu2019}.
Other attacks include Attribute inference~\cite{He2020,Wang2019}, obtaining attributes of clients rather than data, and Membership Inference Attacks~\cite{Chen2020,Ye2022}, which allow the attacker to infer whether an individual was part of the training dataset.

In our experimental setup we assume an honest environment.
Our consortium comprises locations that are familiar with one another and share a common goal of advancing research, while adhering to strict privacy constraints.
In such scenarios FL offers a privacy by design structure during the learning process.
Still, one must be concerned about attacks that can be carried out on the final resulting model weights such as Model Inversion.
In previous work, multiple factors mitigating the possible success of such attacks have been published~\cite{Usynin2021}.
These include knowledge distillation, as the model learns a distribution over the proxy model's output~\cite{Papernot2018}.
Less overfitting causes more general gradients that are less bound to individual samples in the training data set and, thus, complicate reconstruction of input data~\cite{Usynin2021}.
Employing regularization as well as larger batch sizes during training positively influence the privacy guarantees of models~\cite{Kaissis2021}.

Cryptographic methods, such as differential privacy (DP), secure multi-party communication (SMPC), and homomorphic encryption (HE), can further enhance privacy guarantees of federated learning~\cite{Kaissis2020}.
Differential privacy perturbs the gradient update or the input data with zero mean noise equipping each data samples with a plausible deniability of membership in the dataset~\cite{Toelle2022,Abadi2016}.
SMPC protects the model training and update process by distributing computations among multiple parties using secret sharing, while HE encrypts the input data, enabling computations to be performed directly on the encrypted inputs~\cite{Silva2020,Gilad2016}.
However, applying cryptographic methods is out of the scope for this paper and we leave it to future work to investigate its influence.
Further, we assured privacy guarantees with knowledge distillation and less overfitting due to large dataset sizes and regularization.

The model weights of the federated knowledge distilled SWIN-UNETR model are made available as a contribution to open science to enable further research in the cardiac CT imaging on more and diverse downstream tasks.
The federated infrastructure is planned to be re-used for more use cases within the DZHK to enable large-scale AI in cardiovascular research.
Concurrently, more hospitals are joining the federated network.

\section*{Methods}

This manuscript's study and results adhere to all pertinent ethical guidelines and uphold ethical standards in both research conduct and manuscript preparation, in accordance with all relevant laws and regulations concerning human subject treatment. 
Each collaborating site's private retrospective data analysis has received approval from its respective institutional review board. 
Each institutional review board allowed for retrospective data analysis without obtained patient consent since no data is disclosed to any participant in the federation.

\begin{figure}[t]
    \centering
    \includegraphics[width=\textwidth]{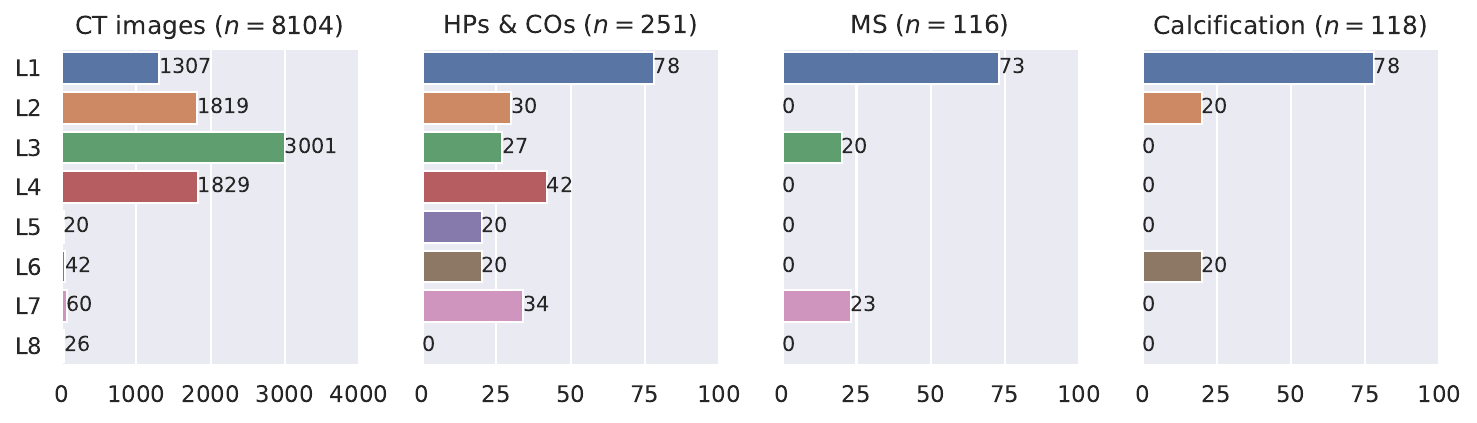}
    \caption{ 
        Data distribution across locations.
        In total $8,104$ CT scans are available across all eight locations. For each label the distribution is differently skewed.
        While the most uniform distribution is present for the hinge point training, for membranous septum and calcification the distribution is skewed. 
        Still, the federated model that is trained over these skewed distributions exhibits better performance than the one trained on a single client. 
        HPs \& COs: Hinge Points and Coronary Ostia Points, MS: Membranous Septum.
    }
    \label{fig6:label-distribution}
\end{figure}

\subsection*{Data}

All procedures performed in studies involving human participants were in accordance with the ethical standards of the institutional and/or national research committee and with the 1964 Helsinki declaration and its later amendments or comparable ethical standards.
This article does not contain any studies with animals performed by any of the authors.
Ethical approval was waived by the local Ethics Committees of Heidelberg (S-475/2021), Göttingen (11/6/21), Hamburg (2021-200262-BO-bet), Munich (21-0497), Münster (2021-487-b-S), Greifswald (BB 091/24), and Frankfurt (2021-366\_1) in view of the retrospective nature of the study and all the procedures being performed were part of the routine care.
In Berlin, a multi-centric study must not explicitly be confirmed when another institutional ethics board waived approval.

This study's data comprises patients who underwent a minimally invasive procedure for replacing their aortic valve with a Transcatheter Aortic Valve Implantation (TAVI) prosthesis.
Consistent with clinical guidelines, every patient undergoes a contrast-enhanced CT scan triggered by an electrocardiogram, conducted in either only the systolic or both the systolic and diastolic phases of the heart cycle.
For this study we included all available contrast enhanced CT scans not dependent whether they only had the systolic or diastolic phase available.
Collective information about the demographics of the included population and CT imaging parameters is presented in Supplementary Figure~1. 

The data acquisition was performed at each participating site from 2015 to 2021. 
Each site's institutional review board approved the retrospective analysis of CT scans from patients who received a TAVI prosthesis during this time.
However, challenges in exporting data from the PACS varied by location, preventing the complete dataset from being utilized for model training or testing at some sites. 
These challenges primarily involved limitations in automatically exporting large volumes of data from the internal PACS systems. 
Our study highlighted deficiencies in data export protocols at some locations, which we hope will trigger investments into better data pipelines. 
Future studies leveraging this infrastructure can benefit from the insights we have gained.

Training is performed on the data quantities across locations as shown in Figure~\ref{fig6:label-distribution}. 
In total, we have 8104 CT images (all locations), 251 hinge points and coronary ostia (HPs \& COs) (7 locations), 116 membranous septum (MS) (3 locations), and 118 calcification labels (3 locations).
None of the displayed data distributions are uniform. 
Location (L) 3 has the highest number of CT images (3001), while L5 has only 20. 
Seven locations have HPs \& COs labeled, with a maximum of 78 cases and a minimum of 20. Three locations contain labels for membranous septum (73/20) and calcification (78/20).
The sample heterogeneity is notably large, especially in comparison to previous studies in the field~\cite{Pati2022,Terrail2023,Linardos2022}. Additionally, no two locations have similar distributions of images or label types.

For each local dataset 20\% of the data was set apart to serve as an independent testset on which to evaluate the final models.
These splits were preserved during the training of all model architectures per subtask as well as for the distilled model version.
We always selected at least one location as test location for each task.
For the hinge points we chose locations 6 and 7 for testing, for membranous septum we chose location 7, and for calcification we again chose location 6.

\subsection*{Harmonized Data Preprocessing}

Subsequent to downloading data in the Digital Imaging and Communications in Medicine (DICOM) file format from the PACS the data was pseudonymized or anonymized dependent on the requirements from the individual institutional review board.
After successful de-identification the data was uploaded in the PACS that is included in the platform.
The platform's filtering and viewing features were utilized to gather the series descriptions of the wanted volumes. 
It is worth noting that there is a significant intra-hospital variance in these descriptions, indicating that they are far from being standardized.
After successful identification we converted DICOMs into the Neuroimaging Informatics Technology Initiative (NIfTI) file format.
This format has the advantage of removing all patient identifying information automatically from the header portion of the DICOM data.
Before performing model training the region containing the heart was focused utilizing the Totalsegmentator tool~\cite{Wasserthal2023}.
Each image was normalized using a CT normalization scheme:

\begin{equation}
    \mathbf{x}_{\mathrm{norm}} = \frac{\mathrm{clip}(\mathbf{X},\mathcal{D}_{0.05},\mathcal{D}_{0.95}) - \mu}{\mathrm{max}(\sigma, 1e-8)} \quad \mathrm{with} \quad \mu=\mathbb{E}[\mathcal{D}] \; \mathrm{and} \; \sigma=\sqrt{\mathbb{V}[\mathcal{D}]} ~,
\end{equation}

\noindent with mean ($\mu=-438.61$) standard deviation($\sigma=520.98$) and the two percentiles ($0.05=-1024$ \& $0.95=696$) were taken from the TotalSegmentator pipeline~\cite{Wasserthal2023}. 

Where not already present, the annotations (3D points for hinge points, origins of coronary arteries, and membranous septum, and segmentations for calcification) were obtained with the medical Medical Interaction Toolkit (MITK)~\cite{Wolf2004}. 
Annotation protocols were provided in text and video form, which was reported to be very beneficial for uniform label generation.

\subsection*{Semi-supervised Federated Knowledge Distillation}

\begin{figure}[t]
    \centering
    \includegraphics[width=0.72\textwidth]{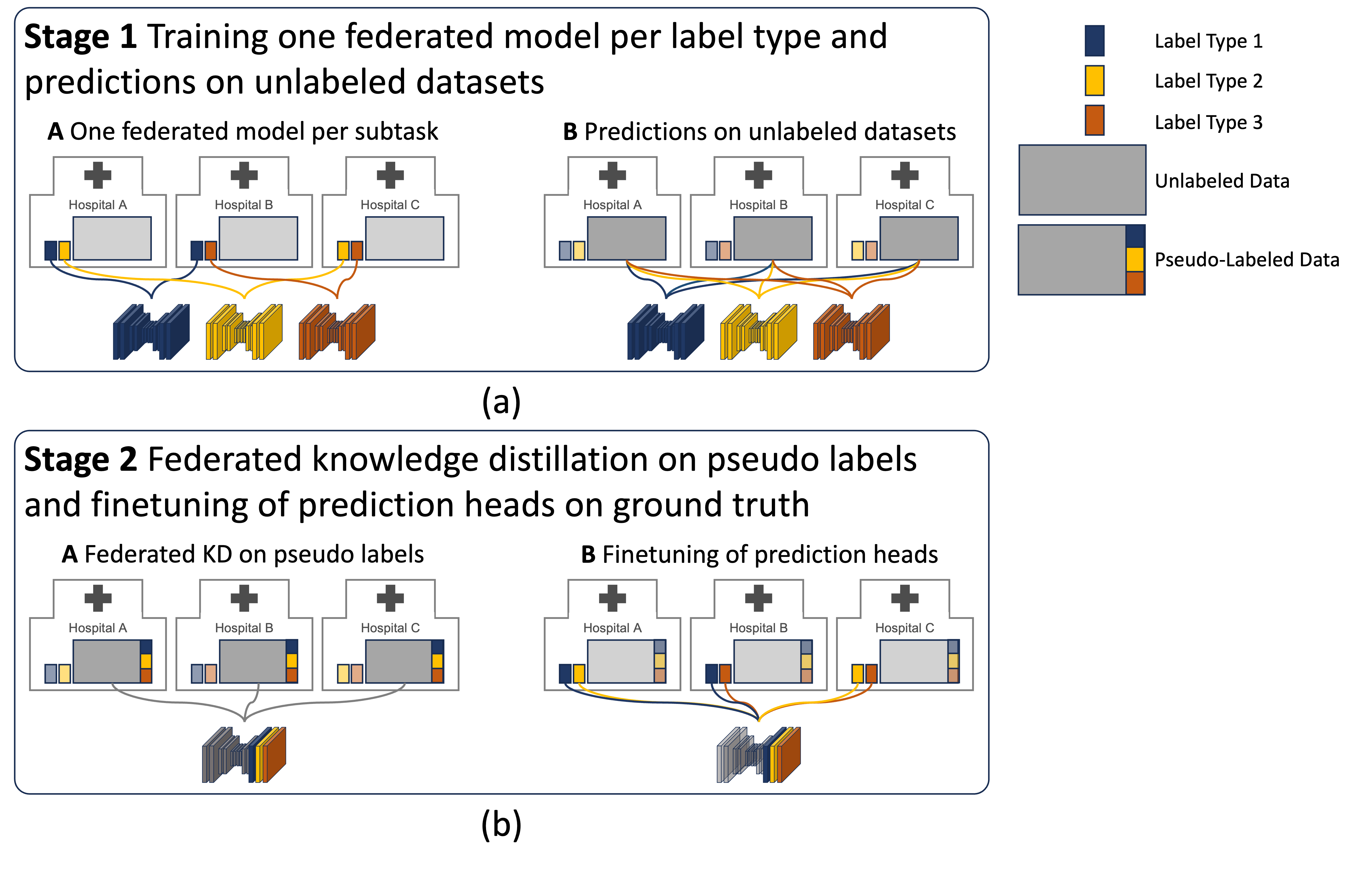}
    \caption{
        Overview of two-stage federated learning process with knowledge distillation.
        a) In \textbf{stage 1} one network per label type is trained with FL across all locations in possession of that particular label type (\textbf{A}).
        These trained models are used to create pseudo labels on the unlabeled datasets residing in each hospital (\textbf{B}).
        b) In \textbf{stage 2}, a transformer based architecture is trained on the generated pseudo labels (\textbf{A}).
        The model gains one prediction head per label type.
        Finally, the model's prediction heads are finetuned on the ground truth labels while leaving the backbone's weights fixed (\textbf{B}).
        Each label type is visualized with a distinct color (blue, yellow, orange), unlabeled data is shown in gray.
        If pseudo-labels are generated for the previously unlabeled data with the models from stage 1, it is marked with the three label colors.
    }
    \label{fig7:method_short}
\end{figure}

Our proposed semi-supervised, two-stage federated learning approach enables effective training on large datasets by leveraging unlabeled data. 
In the first stage, we train a convolutional model on labeled subsets of the data. 
Since some locations only have partial labels, we train a separate model for each label type (i.e., hinge points and coronary arteries, membranous septum, and calcification). 
These three specialized models generate pseudo-labels on the unlabeled data at each site.

In the second stage, we use these pseudo-labels to train a transformer model with a unified structure that includes a prediction head for each label type. 
This setup combines knowledge across the three tasks into one model through a semi-supervised, federated knowledge distillation (KD) process, condensing the knowledge of three models into a single, comprehensive model.
Figure~\ref{fig7:method_short} visualizes the two-stage training procedure.

Finally, we finetune each prediction head on the labeled data at each client, while keeping the weights of the shared feature extractor fixed. 
This design allows the model to learn task-agnostic features in the backbone, meaning features for different segmentation and localization tasks, like identifying the aortic root or the aortic valve hinge points, are captured in a unified manner.

\subsection*{The Neural Network Architectures}

For the three subtasks we used the popular 3D UNet with residual connections (3D-ResUNet) with 32 base filters~\cite{Isensee2021,Cicek2016,He2016}.
The learning rate was set to $lr=0.01$ and optimized with the AdamW optimizer~\cite{Loshchilov2019}.
As loss function during training we used a combination of cross entropy and DICE score loss with deep supervision~\cite{Sudre2017}.
When applying deep supervision also for the intermediate outputs of the skip connections the loss function is applied to a downsampled version of the target, which has been shown to improve segmentation performance~\cite{Isensee2021}.
For guidance we feed the segmented heart obtained from the TotalSegmentator tool~\cite{Wasserthal2023} as a condition such that the models can learn the anatomical relations between heart and the corresponding structures.

For the final model that combines the knowledge from the three subnetworks we use the SWIN UNet Transformer (SWIN-UNETR)~\cite{Hatamizadeh2022b}.
We use a feature size of 24 with a patch size of $\mathbb{R}^{96 \times 96 \times 96}$.
The learning rate was set to $lr=10^{-4}$ and optimized with the AdamW optimizer~\cite{Loshchilov2019}.
We equipped the transformer with three heads, one for each task, to train all tasks concurrently.
We again add the heart segmentation as input for anatomical guidance.
For comparison we also train a conventional vision transformer based segmentation model~\cite{Hatamizadeh2022a}.
We used a hidden dimension of 768 and a patch size of $\mathbb{R}^{16 \times 16 \times 16}$.
Optimizer setting and inputs are similar to the SWIN-UNETR.
The transformer based architectures vary in their attention mechanism.
While the ViT employs conventional self-attention, the SWIN-UNETR uses shifted window self-attention.

\subsection*{The Federation}

In federated learning multiple data holding locations train a model locally on their data shards and report the trained model weights back to a central server where averaging is performed~\cite{McMahan2017,Karimireddy2020}.
After successful averaging another round of training is initiated until the model converges.
Each round is termed a federated round.
This allows data privacy compliant model training as no patient data ever leaves the individual hospitals boundaries.
The most widespread architecture is a hub-and-spoke system were all locations train in parallel instead of an e.g. sequential training~\cite{Kaissis2020,Pati2021}.

Our federation spans eight cardiology and radiology department in university hospitals in Germany (c.f. Figure~\ref{fig1:method}).
Connection could be established only from within the individual clinics to a server that resided behind a firewall at Heidelberg University.
Each model was trained for 20 federated rounds of averaging with 10 local epochs in each round.
We chose to perform model weight's aggregation using a popular variant of the federated averaging algorithm~\cite{Li2020}. 
Every communication in our federation was based on transport layer security, additional authentication with username and password, and server-side IP address white listing.
These measures help mitigate some of the privacy and security concerns still inherent to FL.

Our work covers the whole process of extracting real world data from clinical information systems and subsequent homogenization of data formats across the different sites and label types.
The federated learning software stack was installed at each location that is intended to be used beyond this study for future research.
We created a custom fork of the renowned Kaapana platform~\cite{Scherer2020}.
It allows for a flexible deployment of containerized applications in combination to a picture archiving and communication system (PACS).
To extract the cohorts needed per location we use a custom developed filtering tool~\cite{Toelle2024a}.
Each data type is stored in a custom structured report template such that they can be linked to the corresponding series.
Segmentation objects can also be stored and linked to the referenced image series within the PACS.
\texttt{Fedbiomed} is used as FL library as they provide very sophisticated security measures~\cite{Silva2020}.
All communication is encrypted with transport layer security (TLS) encryption, where the key is distributed to the locations prior to training.
Further, each client must authenticate with custom credentials (username and password).
And last, IP white listing is performed such that only predefined IP addresses can initiate a connection.
The connection is unidirectional.
It must be initiated from within the clinic network, the locations then poll for updates such that no action can be triggered from the server without the client noticing.

\section*{Data Availability}

All data from the eight sites used in this study are not made publicly available due to restrictions imposed by the participating sites.
The data was also not publicly available during conducting of this study.
As by privacy-by-design definition of federated learning they were instead used locally during training and validation of the trained models.
The data to reproduce the plots as well as the corresponding scripts are made publicly available under: \url{https://github.com/Cardio-AI/FedKD-for-Cardiac-CT}.
The ImageCAS dataset is available under: \url{https://github.com/XiaoweiXu/ImageCAS-A-Large-Scale-Dataset-and-Benchmark-for-Coronary-Artery-Segmentation-based-on-CT}.
The corresponding labels for quantifying the inter-observer variability are available at: \url{https://github.com/Cardio-AI/FedKD-for-Cardiac-CT}.
The pointsets can be opened with the Medical Interaction Toolkit (MITK) available under: \url{https://www.mitk.org/wiki/The_Medical_Imaging_Interaction_Toolkit_(MITK)}.

\section*{Code Availability}

Following the FAIR criteria (findability, accessibility, interoperability, and reusability) in scientific research all code used in this study is made publicly available.
We used a custom fork of Kaapana~\cite{Scherer2020} from \url{https://github.com/kaapana/kaapana} which is available under \url{https://github.com/Cardio-AI/kaapana} for orchestration of docker containers at each location.
The federeated learning library \texttt{fedbiomed} is available under \url{https://github.com/fedbiomed/fedbiomed} our custom fork with more security features enabled is avilable under \url{https://github.com/Cardio-AI/fedbiomed}.
For creation of labels we use MITK \url{https://www.mitk.org/wiki/The_Medical_Imaging_Interaction_Toolkit_(MITK)}.
The nnUNet pipeline used for training the per-task models is available under \url{https://github.com/MIC-DKFZ/nnUNet}.
Our preprocessing, training, and validation scripts are made available under \url{https://github.com/Cardio-AI/FedKD-for-Cardiac-CT}.
The pipelines were developed using PyTorch~\cite{Paszke2019}, MONAI~\cite{Cardoso2022}, TorchIO~\cite{Perez-Garcia2021}, and SimpleITK~\cite{Beare2018}.

\section*{Acknowledgements}

The project was funded by the DZHK, the Klaus Tschira Foundation within the Informatics for Life framework, and the BMBF-SWAG Project 01KD2215D.

\section*{Author Contributions}

MT developed the presented method, conducted the experiments and evaluations, set up the federated learning software stack, and wrote the manuscript.
SE organised the consortium, and significantly helped to shape the methods and the manuscript. 
SEb, MB, and HK contributed code for model training and software setup.
PG, LK, NK, AL, SM, CS, JMS, and SS were the direct contact persons at each participating location, set up the local infrastructures, and exported, curated, and uploaded the required data.
FA, PB, GD, NF, SG, AH, AM, EN, SO, MS, TF, TS, and SE developed the idea for the study, provided guidance, and helped with revising the final manuscript.

\section*{Competing interests}

NF reports speaker honoraria, presentations or advisory board consultations from AstraZeneca, Bayer AG, Boehringer Ingelheim, Novartis, Pfizer, Daiichi Sankyo Deutschland.
TS reports research, educational, or travel grants and honoraria for lectures or advisory board consultations from Abbott Vascular, AstraZeneca, BoehringerIngelheim, Bristol Myers Squibb, Corvia, Cytokinetics, Edwards Life Sciences, Medtronic, Myocardia, Novartis, Pfizer, Teleflex.
AM reports consulting or lecturing fees from Medtronic, Bayer, Pfizer.
CS reports speaker honorarium from AstraZeneca.
SE reports speaker honorarium from Boehringer Ingelheim.
None are related to the content of the manuscript.
The authors declare no conflicts of interest.


\newpage

\FloatBarrier
\section*{Supplementary Information}
\FloatBarrier

\renewcommand{\figurename}{Supplementary Figure}
\renewcommand{\tablename}{Supplementary Table}
\setcounter{figure}{0}
\setcounter{table}{0}

\begin{figure}[H]
    \centering
    \includegraphics[width=\textwidth]{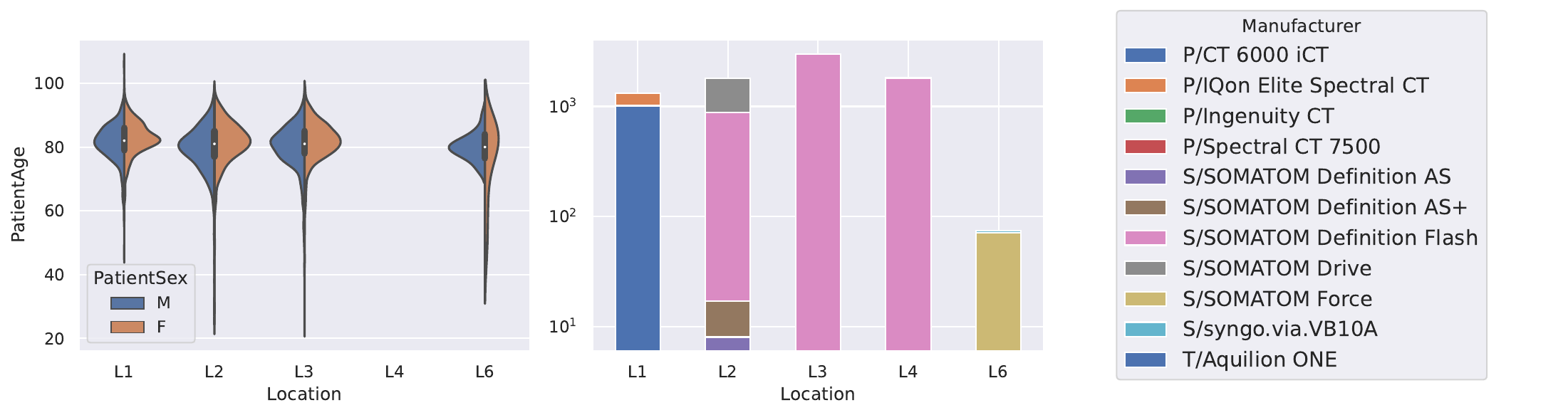}
    \includegraphics[width=\textwidth]{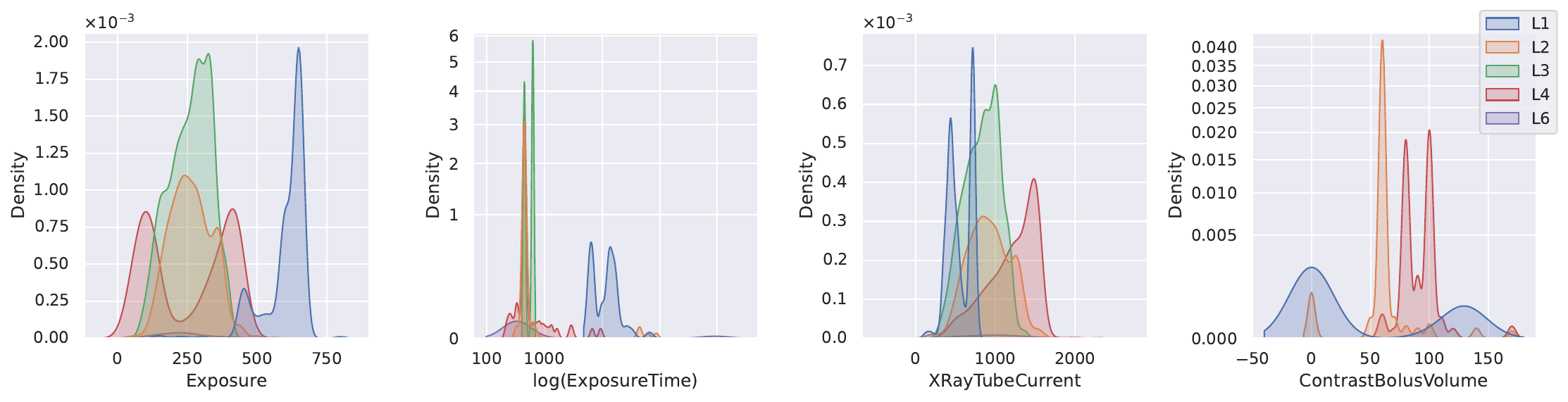}
    \caption{
    Demographics of patients and data properties across locations.
    Some data was not available at all locations. 
    Three manufacturers with in total eleven different models were included in the federated training. The acquisition protocols in terms of exposure, exposure time, X-ray tube current, and contrast bolus volume vary across locations. Manufacture acronyms are P: Philips, S: Siemens, T: Toshiba.
    }
    \label{fig1:demographics}
\end{figure}

\begin{table}[H]
    \centering
    \caption{
    Results of local, federated, and knowledge distilled models per location for the task of detecting hinge points and coronary ostia (HPs \& COs).
    ed and KD are trained on L1,2,3,4,6. The local models often overfit to the training data and even underperform on their respective testset. The federated and especially knowledge distilled models show better generalization. All values are reported in mm with mean and standard deviation.}
    \label{tab:hps_appendix}
    \begin{tabular}{m{.6cm}lccccccccc}
        \toprule
        & Train & L1 & L2 & L3 & L4 & L5 & L6 & L7 \\
        \midrule
        \multirow{7}{*}{\rotatebox{90}{UNet}} &             L1 &    $3.58 \pm 5.17$ &    $3.74 \pm 2.03$ &    $3.67 \pm 1.68$ &    $4.35 \pm 3.94$ &    $3.23 \pm 1.83$ &    $2.85 \pm 1.62$ &    $4.84 \pm 3.46$ \\
        &             L2 &  $17.15 \pm 21.42$ &  $11.65 \pm 12.59$ &  $11.24 \pm 11.08$ &   $12.31 \pm 9.99$ &  $12.63 \pm 12.29$ &  $11.71 \pm 11.43$ &  $13.56 \pm 11.62$ \\
        &             L3 &    $5.98 \pm 7.54$ &     $4.79 \pm 3.2$ &    $5.19 \pm 8.77$ &    $5.01 \pm 3.52$ &    $4.88 \pm 2.17$ &    $3.77 \pm 2.03$ &    $5.35 \pm 3.72$ \\
        &             L4 &    $4.81 \pm 5.35$ &     $4.9 \pm 3.19$ &    $8.0 \pm 10.94$ &    $4.53 \pm 4.05$ &     $4.14 \pm 2.3$ &    $3.81 \pm 5.95$ &    $5.44 \pm 3.84$ \\cardic-ct
        &             L6 &    $4.58 \pm 3.75$ &     $4.83 \pm 3.74$ &    $4.97 \pm 3.22$ &    $4.53 \pm 3.50$ &    $3.29 \pm 4.32$ &    $3.12 \pm 3.96$ &    $4.73 \pm 3.88$ \\
        &            Fed &     $3.92 \pm 5.6$ &    $3.53 \pm 2.36$ &    $3.58 \pm 1.79$ &     $4.12 \pm 4.0$ &    $3.41 \pm 4.18$ &    $2.71 \pm 1.28$ &    $3.86 \pm 3.09$ \\
        &             KD &     $3.84 \pm 5.3$ &    $3.77 \pm 1.93$ &    $3.26 \pm 1.78$ &    $4.43 \pm 4.03$ &    $3.37 \pm 1.83$ &    $2.61 \pm 1.24$ &    $4.59 \pm 3.77$ \\
        \midrule
        \multirow{7}{*}{\rotatebox{90}{ViT}} &             L1 &    $4.93 \pm 5.23$ &     $4.7 \pm 2.48$ &    $3.82 \pm 2.11$ &     $6.1 \pm 3.71$ &     $5.1 \pm 9.43$ &    $3.76 \pm 2.16$ &    $5.75 \pm 3.61$ \\
        &             L2 &  $18.92 \pm 21.33$ &  $15.45 \pm 17.23$ &  $14.63 \pm 13.21$ &   $14.99 \pm 12.15$ &  $13.72 \pm 12.30$ &  $15.78 \pm 17.48$ &  $13.78 \pm 14.63$ \\
        &             L3 &  $15.62 \pm 26.09$ &   $8.65 \pm 14.66$ &    $7.6 \pm 14.63$ &  $13.58 \pm 21.33$ &  $12.68 \pm 20.17$ &   $9.49 \pm 14.03$ &  $16.03 \pm 24.19$ \\
        &             L4 &   $26.87 \pm 7.05$ &   $25.24 \pm 6.42$ &   $24.65 \pm 5.02$ &  $24.75 \pm 10.91$ &   $26.06 \pm 6.18$ &    $23.39 \pm 6.5$ &   $26.33 \pm 5.73$ \\
        &             L6 &    $5.37 \pm 34.83$ &     $5.11 \pm 3.98$ &    $5.12 \pm 4.87$ &    $5.62 \pm 3.28$ &    $4.05 \pm 5.45$ &    $6.36 \pm 4.05$ &    $5.82 \pm 3.96$ \\
        &            Fed &    $4.68 \pm 5.26$ &    $4.47 \pm 2.62$ &    $3.75 \pm 2.18$ &     $5.4 \pm 3.81$ &    $3.87 \pm 1.97$ &    $3.09 \pm 1.28$ &    $5.11 \pm 3.53$ \\
        &             KD &    $4.68 \pm 5.25$ &     $4.4 \pm 2.57$ &    $3.69 \pm 2.14$ &    $5.37 \pm 3.81$ &     $3.9 \pm 1.94$ &    $3.07 \pm 1.42$ &    $5.01 \pm 3.43$ \\
        \midrule
        \multirow{7}{*}{\rotatebox{90}{SWIN}} &             L1 &    $3.61 \pm 6.17$ &    $3.62 \pm 2.22$ &    $3.76 \pm 1.72$ &     $5.8 \pm 7.03$ &    $2.76 \pm 1.47$ &     $2.79 \pm 1.7$ &    $4.29 \pm 3.28$ \\
        &             L2 &  $15.93 \pm 28.61$ &     $3.9 \pm 2.01$ &    $7.02 \pm 8.66$ &   $17.9 \pm 25.69$ &  $15.55 \pm 20.56$ &  $17.13 \pm 12.91$ &  $11.47 \pm 19.15$ \\
        &             L3 &  $15.86 \pm 13.07$ &   $9.99 \pm 10.17$ &    $3.65 \pm 5.66$ &   $14.09 \pm 9.64$ &     $8.2 \pm 9.96$ &    $9.21 \pm 11.0$ &  $10.49 \pm 11.28$ \\
        &             L4 &    $4.66 \pm 7.14$ &    $3.54 \pm 2.27$ &    $3.48 \pm 1.81$ &    $4.84 \pm 6.42$ &    $3.58 \pm 1.83$ &    $1.97 \pm 1.02$ &    $4.17 \pm 3.25$ \\
        &             L6 &    $5.12 \pm 6.23$ &    $3.58 \pm 4.83$ &    $3.21 \pm 2.91$ &    $4.58 \pm 3.27$ &    $2.86 \pm 2.75$ &    $3.42 \pm 2.53$ &    $3.92 \pm 3.33$ \\
        &            Fed &    $4.73 \pm 7.16$ &    $3.47 \pm 1.96$ &    $3.93 \pm 1.85$ &    $5.65 \pm 6.63$ &    $3.02 \pm 1.68$ &    $2.85 \pm 1.49$ &    $4.34 \pm 3.17$ \\
        &             KD &    $3.49 \pm 5.36$ &    $3.26 \pm 1.93$ &    $3.18 \pm 1.78$ &    $4.17 \pm 3.94$ &    $2.94 \pm 1.83$ &    $2.39 \pm 1.11$ &    $4.11 \pm 3.35$ \\
        \bottomrule
    \end{tabular}
\end{table}

\begin{table}[H]
    \centering
    \caption{
    Results of local, federated, and knowledge distilled models per location for the task of detecting the membranous septum (MS).
    Fed and KD are trained on L1 and L3. The local models sometimes overfit to the training data and even underperform on their respective testset. The federated and especially knowledge distilled models show better generalization. All values are reported in mm with mean and standard deviation.}
    \label{tab:ms_appendix}
    \begin{tabular}{m{.8cm}m{.8cm}m{3cm}<{\centering}m{3cm}<{\centering}m{3cm}<{\centering}}
        \toprule
        & Train & L1 & L3 & L7 \\
        \midrule
        \multirow{4}{*}{\rotatebox{90}{UNet}}   & L1  & $3.45 \pm 2.63$ & $5.10 \pm 0.60$ & $5.01 \pm 2.33$ \\
        & L3  & $4.68 \pm 2.73$ & $3.66 \pm 1.06$ & $4.36 \pm 1.88$ \\
        & Fed & $4.64 \pm 2.33$ & $3.72 \pm 1.34$ & $4.37 \pm 2.41$ \\
        & KD  & $3.26 \pm 2.34$ & $3.25 \pm 1.32$ & $3.40 \pm 1.56$ \\
        \midrule
        \multirow{4}{*}{\rotatebox{90}{ViT}}  & L1  & $3.55 \pm 2.55$ & $3.29 \pm 1.53$ & $4.26 \pm 2.65$ \\
        & L3  & $54.28 \pm 36.64$ & $24.52 \pm 18.99$ & $53.98 \pm 34.95$ \\
        & Fed & $3.69 \pm 2.54$ & $4.49 \pm 1.91$ & $5.39 \pm 2.64$ \\
        & KD  & $3.34 \pm 2.39$ & $2.97 \pm 1.50$ & $3.60 \pm 1.56$ \\
        \midrule
        \multirow{4}{*}{\rotatebox{90}{SWIN}}   & L1  & $4.44 \pm 3.55$ & $4.75 \pm 1.98$ & $4.92 \pm 1.63$ \\
        & L3  & $3.94 \pm 2.33$ & $3.04 \pm 0.91$ & $4.60 \pm 2.31$ \\
        & Fed & $3.17 \pm 2.43$ & $3.30 \pm 1.60$ & $3.43 \pm 1.44$ \\
        & KD  & $3.29 \pm 2.44$ & $2.72 \pm 0.96$ & $3.29 \pm 1.45$ \\
        \bottomrule
    \end{tabular}
\end{table}

\begin{table}[H]
    \centering
    \caption{
    Results of local, federated, and knowledge distilled models per location for the task of segmenting the calcification.
    Fed and KD are trained on L1 and L2. The DICE scores are reported with mean and standard deviation.}
    \label{tab:calc_appendix}
    \begin{tabular}{m{.8cm}m{.8cm}m{3cm}<{\centering}m{3cm}<{\centering}m{3cm}<{\centering}}
    \toprule
    & Train & L1 & L2 & L6 \\
    \midrule
    \multirow{4}{*}{\rotatebox{90}{UNet}}   & L1  & $0.593 \pm 0.233$ & $0.539 \pm 0.134$ & $0.583 \pm 0.412$ \\
    & L2  & $0.391 \pm 0.170$ & $0.401 \pm 0.207$ & $0.272 \pm 0.274$ \\
    & Fed & $0.486 \pm 0.193$ & $0.515 \pm 0.246$ & $0.391 \pm 0.212$ \\
    & KD  & $0.537 \pm 0.177$ & $0.500 \pm 0.275$ & $0.526 \pm 0.228$ \\
    \midrule
    \multirow{4}{*}{\rotatebox{90}{ViT}}  & L1  & $0.694 \pm 0.136$ & $0.616 \pm 0.268$ & $0.663 \pm 0.241$ \\
    & L2  & $0.378 \pm 0.129$ & $0.516 \pm 0.209$ & $0.327 \pm 0.272$ \\
    & Fed & $0.680 \pm 0.138$ & $0.648 \pm 0.272$ & $0.636 \pm 0.274$ \\
    & KD  & $0.569 \pm 0.169$ & $0.542 \pm 0.248$ & $0.566 \pm 0.231$ \\
    \midrule
    \multirow{4}{*}{\rotatebox{90}{SWIN}}   & L1  & $0.704 \pm 0.138$ & $0.647 \pm 0.285$ & $0.661 \pm 0.243$ \\
    & L2  & $0.384 \pm 0.199$ & $0.519 \pm 0.236$ & $0.312 \pm 0.222$ \\
    & Fed & $0.667 \pm 0.155$ & $0.652 \pm 0.277$ & $0.682 \pm 0.230$ \\
    & KD  & $0.652 \pm 0.176$ & $0.627 \pm 0.273$ & $0.670 \pm 0.231$ \\
    \bottomrule
    \end{tabular}
\end{table}

\end{document}